\documentclass[showpacs,preprintnumbers,aps,prd,eqsecnum,amsmath,amssymb,amsfonts]{revtex4}
\usepackage[dvips]{graphicx}
\usepackage{latexsym}
\usepackage{bm}

\begin{document}

\title{
Primordial gravitational waves in inflationary braneworld
}

\author{Tsutomu Kobayashi}
\email{tsutomu@tap.scphys.kyoto-u.ac.jp}
\author{Hideaki Kudoh}
\email{kudoh@yukawa.kyoto-u.ac.jp}
\author{Takahiro Tanaka${^a}$} 
\email{tanaka@yukawa.kyoto-u.ac.jp}

\affiliation{
Department of Physics, Kyoto University, Kyoto 606-8502, Japan 
\\
${^a}$
Yukawa Institute for Theoretical Physics, 
Kyoto University, Kyoto 606-8502, Japan
}

\begin{abstract}
We study primordial gravitational waves from inflation in the Randall-Sundrum braneworld model.
The effect of small change of the Hubble parameter during inflation 
is investigated using a toy model given by 
connecting two de Sitter branes. 
We analyze the power spectrum of the final zero-mode gravitons,   
which is generated from the vacuum fluctuations of 
both the initial Kaluza-Klein modes and the zero mode.
The amplitude of fluctuations is confirmed to agree with the four-dimensional
one at low energies, whereas it is enhanced due to the normalization
factor of the zero mode at high energies.
We show that the five-dimensional spectrum can be well 
approximated by applying a simple mapping 
to the four-dimensional fluctuation amplitude. 
\end{abstract}

\pacs{04.50.+h, 11.10.Kk, 98.80.Cq}

\preprint{KUNS-1842, YITP-03-23}

\maketitle

\section{Introduction}

Our four-dimensional universe might be embedded as a
three-brane in a higher dimensional spacetime with all the standard model
particles confined to the brane and with gravity allowed to propagate in the
extra dimensions. 
This so-called braneworld scenario opened various possibilities. 
A possible solution to the hierarchy problem in particle physics was
presented by introducing large extra dimensions \cite{Arkani-Hamed} or 
a new type of compactification \cite{RS1}. 
Also interesting is a new possibility 
proposed in Ref.~\cite{RS2} that  
the four-dimensional gravity is recovered 
effectively on the brane despite the infinite extension of the extra dimension
\cite{GarrigaTanaka,Kudoh:2001wb}.

The braneworld scenarios have also had a large impact on cosmology
(for a review of cosmological aspects of the braneworld scenarios, see,
e.g., \cite{Langlois:review}).
Homogeneous and isotropic cosmological models have been built
\cite{brane_cosmology}, and various types of inflation models
proposed \cite{GarrigaSasaki, brane_inf, brane_inf_2, dens, Kobayashi:2000yh}.
A lot of effort has been put forth in searching for new and
characteristic features of the braneworld cosmology. For example,
cosmological perturbations and the related physics of the early universe
have attracted large attention in the expectation that the
braneworld inflation might leave their characteristic prints on the
primordial spectrum of perturbations.

Although the cosmological perturbations have been discussed in a number of
publications (see, e.g.,~\cite{pert, Koyama:2000cc, dens, Lan1, Rubakov,
Frolov:2002qm, Langlois:review, Deruelle:2002cd, Kobayashi:2000yh, Sahni} and references
therein), the presence of the extra dimension does not allow
detailed predictions about the cosmological consequences. 
As the simplest case, the primordial spectra of density perturbations
and of gravitational (tensor) perturbations were investigated 
neglecting the nontrivial evolution of perturbations in the
bulk in a model of slow-roll inflation driven by a inflaton 
field confined to the brane~\cite{dens,Lan1,Sahni}.
(We must mention that gravitational waves have been considered in the
context of braneworld models other than the Randall-Sundrum type as
well; see \cite{Koyama:2003yz, Giudice:2002vh}.)
Although complicated, the effect on the scalar type perturbations 
due to the evolution of perturbations in the bulk 
has been discussed under some assumptions~\cite{Koyama:2000cc}. 
As for the gravitational wave perturbations, 
the authors of Ref.~\cite{Rubakov} 
considered a simplified inflation model in which de Sitter stage of
inflation is instantaneously connected to Minkowski space.  
In this model it is possible to 
solve the perturbation equations including the bulk to some extent. 
They focused on perturbations with comoving scale 
exceeding the Hubble scale at the end of inflation.

Also in this paper we study gravitational waves from an inflating brane. 
If the Hubble parameter on the brane is constant, the power
spectrum becomes scale-invariant~\cite{Lan1}.
However, the Hubble parameter usually changes even during inflation.  
The change of the Hubble parameter, i.e., the nontrivial motion of the brane in 
the five-dimensional bulk, ``disturbs'' the graviton wave function. 
As a result, zero-mode gravitons, which correspond to the
four-dimensional gravitational waves, are created from vacuum
fluctuations in the Kaluza-Klein modes as well as in the zero mode. 
It is also possible that gravitons initially in the zero mode 
escape into the extra dimension as `dark radiation'. 
Therefore we expect that the nontrivial motion of the brane may
leave characteristic features of the braneworld inflation.  
If so, it is
interesting to search for a signature of the extra dimension left on
the primordial spectrum.
However, there is a technical difficulty. 
When the Hubble parameter is time dependent,
the bulk equations are no longer separable.  
Then we have to solve a complicated
partial differential equation.
To cope with this difficulty, we consider a simple model in which 
two de Sitter branes are joined at a certain time;
namely, we assume
that the Hubble parameter changes discontinuously. 
In this model we can calculate the 
power spectrum almost analytically.
This is a milder version of the transition described in \cite{Rubakov}.

This paper is organized as follows.
In the next section we describe the setup of our five-dimensional model, 
and explain the formalism introduced in Ref.~\cite{Rubakov} to solve the mode 
functions for gravitational wave perturbations. 
Using this formalism, 
we explicitly evaluate the Bogoliubov coefficients in Section 3.
In Section 4 we translate the results for the Bogoliubov coefficients into 
the power spectrum of gravitational waves, and its properties are discussed. 
We show that the power spectrum for our five-dimensional 
model can be reproduced with good accuracy from that 
for the corresponding four-dimensional model by applying 
a simple mapping. 
Section 5 is devoted to conclusion.

\section{de Sitter brane and gravitational wave perturbations}

\subsection{Background metric, gravitational wave perturbations, 
and mode functions \label{background, per}}

Let us start with the simple case in which the background is 
given by a pure de Sitter brane in AdS$_5$ bulk spacetime.
We solve the five-dimensional Einstein equations 
for gravitational wave perturbations. 
For this purpose, 
it is convenient to use a coordinate system in which the position of the
brane becomes a constant coordinate surface. 
In such a coordinate system the background metric is written as 
    \begin{eqnarray}
    ds^2 = \frac{\ell^2}{(\sinh \xi)^2} \left[
    \frac{1}{\eta^2} ( - d \eta^2 + \delta_{ij}
    dx^i dx^j ) + d \xi^2 \right],
    \label{dSBrane_coordi}
    \end{eqnarray}
where $\ell$ is the bulk curvature radius, and 
the de Sitter brane is placed at $\xi =$ const $= \xi_B$.
Note that here $\eta$ is supposed to be negative. 
On the brane, the scale factor is given by $a(\eta) = 1/(-\eta H)$ 
and the Hubble parameter becomes 
\begin{eqnarray}
H = \ell^{-1} \sinh \xi_B. 
\end{eqnarray}
Note that under the coordinate transformations
\begin{eqnarray}
    t &=& \eta \cosh \xi - \eta_0 \cosh \xi_B,
\nonumber
\\
    z &=& - \eta \sinh \xi,
\end{eqnarray}
with a constant $\eta_0$, the metric (\ref{dSBrane_coordi}) becomes 
the AdS$_5$ metric in the Poincar\'{e} coordinates.

The metric with gravitational wave perturbations is written as 
    \begin{eqnarray}
    ds^2 = \frac{\ell^2}{(\sinh \xi)^2} \left\{
    \frac{1}{\eta^2} \left[ - d \eta^2 + \left( \delta_{ij}
    + h_{ij}^{{\rm TT}} \right)
    dx^i dx^j \right] + d \xi^2 \right\}. 
    \end{eqnarray}
We decompose the transverse-traceless tensor $h_{ij}^{{\rm TT}}$ 
into the spatial Fourier modes as 
    \begin{eqnarray}
    h_{ij}^{{\rm TT}}(\eta, \bm{x}, \xi) = 
    \frac{\sqrt{2}}{(M_5)^{3/2}} \cdot
    \frac{1}{( 2 \pi )^{3/2}}
    \int d^3 p~ \phi(\eta, \xi ;p)
    e^{i \bm{p} \cdot \bm{x}} e_{ij}, 
    \label{Fourier}
    \end{eqnarray}
where $e_{ij}$ is the polarization tensor, and the summation over
different polarization was suppressed. 
$M_5$ represents the five-dimensional Planck mass, and it 
is related to the four-dimensional Planck mass $M_{{\rm Pl}}$ by $\ell M_5^3 = M_{{\rm Pl}}^2$.
The factor $\sqrt{2}/(M_5)^{3/2}$ is chosen so that the effective
action for $\phi$ corresponds to the action for the canonically
normalized scalar field.
Then, the Einstein equations for the gravitational wave perturbations 
reduce to the Klein-Gordon equation for a massless scalar field in AdS$_5$,
    \begin{eqnarray}
    \Box \phi =
    \left( {\cal D}_{\eta} - {\cal D}_{\xi} \right) \phi = 0,
    \label{KG_in_dS} 
    \end{eqnarray}
where the derivative operators are defined by
    \begin{eqnarray}
    {\cal D}_{\eta} = \eta^2 \frac{\partial^2}{\partial \eta^2}
    - 2 \eta \frac{\partial}{\partial \eta}+ p^2 \eta^2,~~
    \nonumber
    \\
    {\cal D}_{\xi} = (\sinh \xi)^3 \frac{\partial}{\partial \xi}
    (\sinh \xi)^{-3}\frac{\partial}{\partial \xi}.
    \end{eqnarray}
We assume $Z_2$ symmetry across the brane.
Assuming that anisotropic stress is zero on the brane, Israel's junction condition gives the boundary condition for the perturbations as
    \begin{eqnarray}
    \left. \partial_{\xi} \phi \right|_{\xi= \xi_B} = 0.
    \label{BC}
    \end{eqnarray}

Since the equation is separable, 
the mode functions are found in the form of 
$\phi_{\kappa} (\eta, \xi) = \psi_{\kappa} (\eta) \chi_{\kappa}
(\xi)$, where $\psi_{\kappa} (\eta)$ and $\chi_{\kappa} (\xi)$ 
satisfy 
    \begin{eqnarray}
    \left({\cal D}_{\eta} + \kappa^2 + \frac{9}{4} \right)
    \psi_{\kappa}(\eta) = 0,
    \label{sep_time_KK}
    \\
    \left({\cal D}_{\xi} + \kappa^2 +\frac{9}{4} \right)
    \chi_{\kappa}(\xi) = 0,
    \label{sep_space_KK}
    \end{eqnarray}
respectively, and the separation constant $\kappa ~(\geq 0)$ is related
to the Kaluza-Klein mass $m$ as $m^2 = (\kappa^2 +9/4) H^2$. 
For $\kappa^2=-9/4$, we have one discrete mode, which is called 
the zero mode. 
The zero-mode wave function is 
given by 
    \begin{eqnarray}
    \phi_{0}^{(\pm)} =  \ell^{-1/2} C(H)
    \frac{H}{\sqrt{2 p}} e^{\pm ip \eta}
    \left( \eta \pm \frac{i}{p} \right),
    \label{zeromode}
    \end{eqnarray}
which is independent of $\xi$. 
The factor $C(H)$ is to be determined by 
the normalization condition, $ ( \phi_0^{(\pm)} \cdot \phi_0^{(\pm)} ) = \mp 1$,
with respect to the Klein-Gordon inner product 
    \begin{eqnarray}
    (F \cdot G) := -2i \int^{\infty}_{\xi_B}
    \frac{\ell^3 d\xi}{(\sinh \xi)^3 \eta^2} \left(
    F \partial_{\eta}G^* -
    G^* \partial_{\eta}F \right). 
    \label{inner_product}
    \end{eqnarray}
Then we have
    \begin{eqnarray}
    C^2 (H) 
    &=&
    \left[ 2 (\sinh \xi_B )^2 \int^{\infty}_{\xi_B}
    d \xi \frac{1}{(\sinh \xi )^3} \right]^{-1}
    \nonumber\\
    &=& \left[
    \sqrt{1 + \ell^2 H^2} + \ell^2 H^2
    \ln \left(
    \frac{\ell H}{1 + \sqrt{1 + \ell^2 H^2}}
    \right)
    \right]^{-1}.
    \label{normalization;C(H)}
    \end{eqnarray}
This normalization factor is the same that was introduced in, 
for example, Refs.~\cite{Frolov:2002qm, Lan1}, and behaves like
    \begin{eqnarray}
    C^2 (H)
    \sim \left\{
    {\displaystyle
    1,
    ~~~~~~(\mbox{at low energies}~~\ell H \ll 1),
    }
    \atop {\displaystyle
    \frac{3}{2}\ell H,
    ~~~~(\mbox{at high energies}~~\ell H \gg 1). 
    }\right.
    \end{eqnarray}

The continuous spectrum called Kaluza-Klein (KK) modes 
starts with $\kappa=0$. 
Notice that the mode labeled by $\kappa =0$ does not correspond to the 
zero mode~\cite{GarrigaSasaki}. 
Writing the positive and the negative frequency modes,  
respectively, as $\phi_{\kappa}^{(+)} =
\psi_{\kappa}^{(+)} \chi_{\kappa}$ and 
$\phi_{\kappa}^{(-)}= (\phi_{\kappa}^{(+)})^* =  \psi_{\kappa}^{(-)}
\chi_{\kappa}^*$, we impose the conditions 
    \begin{eqnarray}
    \frac{i\ell^3}{\eta^2} \left(
    \psi^{(+)}_{\kappa} \partial_{\eta}
    \psi^{(-)}_{\kappa} - c.c. \right) = 1,
    \\
    2 \int^{\infty}_{\xi_B} \frac{d\xi}{(\sinh \xi)^3}
    \chi_{\kappa'}^* \chi_{\kappa}
    = \delta(\kappa - \kappa'),
    \label{normal_chi}
    \end{eqnarray}
so as to satisfy the normalization condition $(\phi_{\kappa}^{(\pm)} \cdot
\phi_{\kappa'}^{(\pm)} ) = \mp \delta ( \kappa - \kappa')$.  The
solutions of Eq.~(\ref{sep_time_KK}) are given in terms of the Hankel
functions by 
    \begin{eqnarray}
    \psi_{\kappa}^{(-)}(\eta) &=&
    \frac{\sqrt{\pi}}{2} \ell^{-3/2}
    e^{- \pi \kappa/2}|\eta|^{3/2}
    H^{(1)}_{i\kappa}(p|\eta|). 
    \end{eqnarray}
The spatial mode function $\chi_{\kappa}(\xi)$ is given in terms of the
associated Legendre functions by \cite{Kobayashi:2000yh}
    \begin{eqnarray}
    \chi_{\kappa} = C_1 (\sinh \xi)^2 \left[
    P_{-1/2 + i \kappa}^{-2} (\cosh \xi)
    - C_2 Q_{-1/2 + i \kappa}^{-2} (\cosh \xi)
    \right],  
    \end{eqnarray}
where from Eqs.~(\ref{BC}) and (\ref{normal_chi})
the  constants $C_1$ and $C_2$ are 
     \begin{eqnarray}
    C_1 &=& \left[ \left|\frac{\Gamma(i \kappa)}{\Gamma(5/2+i \kappa)}\right|^2
    + \left| \frac{\Gamma(-i \kappa)}{\Gamma(5/2-i \kappa)}- \pi C_2
    \frac{\Gamma(i \kappa-3/2)}{\Gamma(1+i \kappa)} \right|^2
    \right]^{-1/2},
    \nonumber
    \\
    C_2 &=& \frac{P^{-1}_{-1/2+i\kappa}(\cosh \xi_B)}
    {Q^{-1}_{-1/2+i\kappa}(\cosh \xi_B)}.
    \end{eqnarray}

As will be seen, we need to evaluate the value of the wave function at the location 
of the brane
$\chi_{\kappa}(\xi_B)$, and in some special cases $\chi_{\kappa}(\xi_B)$
reduces to a rather simple form. 
For $\sinh \xi_B \ll 1$ \textit{and} $\kappa \sinh \xi_B \ll 1$, we have 
    \begin{eqnarray}
    \chi_{\kappa}(\xi_B) \approx
    \sqrt{\frac{\kappa \tanh \pi \kappa}{2}}
    \sqrt{
    \frac{\kappa^2 + 1/4}{\kappa^2 + 9/4}
    } (\sinh \xi_B)^2,
    \label{H_LL_1_and_kH_LL_1}
    \end{eqnarray}
while, for $\sinh \xi_B \gg 1$ \textit{or} $\kappa \sinh \xi_B \gg 1$, we have 
    \begin{eqnarray}
    \chi_{\kappa}(\xi_B) \approx \frac{1}{\sqrt{\pi}}
    ( \sinh \xi_B )^{3/2} \frac{\kappa}
    {\sqrt{\kappa^2 + 9/4}}.
    \label{H_GG_1_or_kH_GG_1}
    \end{eqnarray}
For the derivation of these two expressions, see Ref.~\cite{Rubakov}.

\subsection{Model with a jump in the Hubble parameter}

We consider a model in which the Hubble parameter changes during inflation.
As we have explained, in the case of constant Hubble parameter 
the brane can be placed at a constant coordinate plane.
When the Hubble parameter varies, we need to consider 
a moving brane in the same coordinates.
For simplicity, we consider the situation in which 
the Hubble parameter changes discontinuously at $\eta = \eta_0$ from $H_1$ to 
\begin{eqnarray}
    H_2 = H_1 - \delta H. 
\end{eqnarray}
Here $\delta H / H_1$ is assumed to be small.
For later convenience, we define a small quantity $\epsilon_H$ by
    \begin{eqnarray}
    \epsilon_H & = & \frac{\ell H_1
    \sqrt{1 + (\ell H_2)^2} - \ell H_2
    \sqrt{1 + (\ell H_1)^2}}
    {\ell H_2}\cr
    & = &  \frac{1}
    {\sqrt{1 + ( \ell H_1 )^2}}
    \frac{\delta H}{H_1} +
    \frac{2 + 3 (\ell H_1)^2}
    {2 (1 + (\ell H_1)^2)^{3/2}}
    \left( \frac{\delta H}{H_1} \right)^2
    + {\cal{O}} \left( \frac{\delta H}{H_1} \right)^3 .
    \end{eqnarray}

\begin{figure}[htbp]
  \begin{center}
   \includegraphics[keepaspectratio=true,height=60mm]{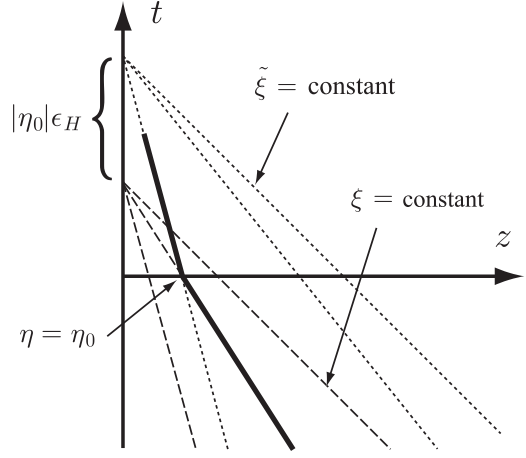}
  \end{center}
  \caption{Trajectory of the brane (thick solid line) in static coordinates.
  Dashed (respectively dotted) lines represent surfaces
  of $\xi =$ constant (respectively $\tilde{\xi}=$ constant).}
  \label{fig:coordi.eps}
\end{figure}

To describe the motion of the de Sitter brane after transition, 
it is natural to introduce a new coordinate system 
$(\tilde{\eta}, \tilde{\xi})$ defined by 
    \begin{eqnarray}
    t 
    &=& \tilde{\eta} \cosh \tilde{\xi}-
    \tilde{\eta}_0 \cosh \tilde{\xi}_B,
    \cr
    z 
    &=& -\tilde{\eta} \sinh \tilde{\xi} .
    \end{eqnarray}
Then, the brane expanding with Hubble parameter $H_2$ is placed 
at $\tilde\xi=\tilde{\xi}_B$ by choosing two constants $\tilde\xi_B$ and 
$\tilde\eta_0$ so as to satisfy $H_2 = \ell^{-1} \sinh \tilde{\xi}_B$ 
and $\eta_0\sinh\xi_B=\tilde\eta_0\sinh\tilde\xi_B$. 
The trajectory of the brane is shown in Fig.~\ref{fig:coordi.eps}.
Apparently, mode functions in this coordinate system take the same form
as those in the previous section, but the arguments 
$({\xi}, {\eta})$ and the Hubble parameter $H_1$ 
are replaced by $(\tilde{\xi}, \tilde{\eta})$ and $H_2$.
We refer to these second set of modes as $\tilde{\phi}_0$ and $\tilde{\phi}_{\kappa}$.
The relation between $(\eta, \xi)$ and $(\tilde{\eta}, \tilde{\xi})$ is
\begin{eqnarray}
\tilde{\eta} &=& - \sqrt{\eta^2 + 2 \epsilon_H \eta_0 \eta \cosh \xi +\epsilon_H^2 \eta_0^2},
\cr
\tanh \tilde{\xi} &=& (\eta \cosh \xi + \epsilon_H \eta_0)^{-1} \eta \sinh \xi.
\label{tilde eta xi}
\end{eqnarray}

As explained above, the variation of the Hubble parameter is assumed to be small.
For a technical reason, we further impose a weak restriction 
that the wavelength of the perturbations concerned is larger than $\delta H/H^2$. 
These conditions are summarized as follows:
\begin{eqnarray}
\frac{\delta H}{H} \ll 1 \,,
\label{assumption 1}
\\
p | \eta_0 | \frac{\delta H}{H} \ll 1 \, .
\label{assumption 2}
\end{eqnarray}

\subsection{Method to calculate Bogoliubov coefficients}

We consider the graviton wave function $\varphi$ that becomes the
zero mode $\tilde{\phi}_0^{(-)}$ at the infinite future $\tilde{\eta}
= 0$.
We write the wave function $\varphi$ as 
    \begin{eqnarray}
    \varphi = \tilde{\phi}_0^{(-)} + \delta \varphi, 
    \label{del phi pre}
    \end{eqnarray}
where the second term $\delta \varphi$ arises because
$\tilde\phi_0^{(-)}$ does not satisfy the boundary condition for
$t<0$. 
Writing down $\tilde{\phi}_0^{(-)}$ and $\delta \varphi$ at the
infinite past, 
$\eta = - \infty$, as a linear combination of $\phi_0^{(\pm)}$ and
$\phi_{\kappa}^{(\pm)}$, we can read the Bogoliubov coefficients. 

At $\eta\to -\infty$ the first term $\tilde{\phi}_0^{(-)}$ is expanded as 
    \begin{eqnarray}
    \tilde{\phi}_{0}^{(-)} \mathop{\longrightarrow}_{\eta\to -\infty} 
    \sum_{ M=0, \kappa }
    \left(
    U_{0 M} \phi_{M}^{(-)} +
    V_{0 M} \phi_{M}^{(+)}
    \right),
    \label{del phi}
    \end{eqnarray}
where the summation is taken over the zero mode and the KK modes.
The coefficients $U_{0\kappa}$ and $V_{0\kappa}$ are written by using  
the inner product as 
    \begin{eqnarray}
    U_{0 M} &=& \lim_{\eta\to -\infty}\left( \tilde{\phi}_{0}^{(-)}
    \cdot \phi_M^{(-)} \right),
    \nonumber
    \\
    V_{0 M} &=& -\lim_{\eta\to -\infty} \left( \tilde{\phi}_{0}^{(-)}
    \cdot \phi_M^{(+)} \right).
    \label{innerp}
    \end{eqnarray}
Evaluation of the inner product at 
$\eta = - \infty$ leads to $V_{00} = V_{0 \kappa} = 0$ \cite{Rubakov},
    while 
    \begin{eqnarray}
    U_{0 0} &\approx& \frac{H_2 C(H_2)}
    {H_1 C (H_1)} e^{
    - i\epsilon_H p \eta_0 \cosh \xi_B}
    \left[ 1 - i \epsilon_H p \eta_0
    \left(
    C^2 (H_1) - \cosh \xi_B
    \right)
    - \frac{1}{2}( \epsilon_H p \eta_0 )^2
    ( \sinh \xi_B )^2 \right],
    \label{U00}
\\
    U_{0 \kappa} &\approx&
    \epsilon_H p \eta_0 \frac{-2 i e^{i \pi /4}}
    {\kappa^2 + 1/4} \frac{\ell H_2}
    {( \ell H_1 )^2} C ( H_2 )
    \chi_{\kappa}( \xi_B ).
    \label{U0kappa}
    \end{eqnarray}
These expressions are approximately correct as long as $p|\eta_0| \delta H / H_1 \ll 1$ is satisfied.
The derivation of these equations is explained in Appendix \ref{app:inner pro}.

The second term $\delta \varphi$ in Eq.~(\ref{del phi pre}) is obtained as follows.
Since both $\varphi$ and $\tilde{\phi}_{0}^{(-)}$ satisfy the
Klein-Gordon equation in the bulk (\ref{KG_in_dS}), $\delta \varphi$ also obeys the
same equation. 
The boundary condition for $\delta \varphi$ is derived from 
(\ref{BC}) as 
$\partial_{\xi} \delta \varphi = - \partial_{\xi}
\tilde{\phi}_{0}^{(-)}$.  
Therefore, the solution $\delta \varphi$ is given by
    \begin{eqnarray}
    \delta \varphi = 2 \int^{\eta_0}_{-\infty} d \eta'
    G_{{\rm adv}}( \eta, \xi ; \eta', \xi_B ) \left[
    \partial_{\xi'}
    \tilde{\phi}_{0}^{(-)}( \eta', \xi' )
    \right]_{\xi'=\xi_B},
    \label{sol}
    \end{eqnarray}
with the aid of the 
advanced Green's function that satisfies
    \begin{eqnarray}
    ( {\cal D}_{\eta} - {\cal D}_{\xi} )
    G_{{\rm adv}}(\eta, \xi; \eta', \xi') =
    \delta( \eta - \eta') \delta( \xi - \xi'). 
    \end{eqnarray}
The explicit form of the Green's function is \cite{Rubakov}
    \begin{eqnarray}
    G_{{\rm adv}} ( \eta, \xi ; \eta' , \xi' )
    = \sum_{M} \frac{i \ell^3}
    {(\sinh \xi')^3 {\eta'}^4} \theta
    ( \eta' - \eta ) \left[ \phi_{M}^{(+)}
    ( \eta, \xi ) \phi_{M }^{(-)} ( \eta', \xi' )
    - \phi_{M }^{(-)}( \eta, \xi )
    \phi_{M }^{(+)}( \eta', \xi' ) \right].
    \label{Green}
    \end{eqnarray}
Taking the limit $\eta \to - \infty$,  we can expand $\delta \varphi$ in terms of in-vacuum mode functions, 
    \begin{eqnarray}
    \delta \varphi \mathop{\longrightarrow}_{\eta\to -\infty} \sum_{M=0, \kappa}
    \left[ u_{0 M } \phi_{M}^{(-)}+
      v_{0 M} \phi_{M}^{(+)}  \right],
    \label{del phi = mode sum}
    \end{eqnarray}
where the coefficients are given by 
    \begin{eqnarray}
    u_{0 M} &=& -2i\ell^3 \int^{\eta_0}_{-\infty}
    \frac{d\eta}{(\sinh \xi_B)^3 {\eta}^4}
    \phi_M^{(+)}(\eta, \xi_B) \left[
    \partial_{\xi} \tilde{\phi}_{0}^{(-)}(\eta, \xi)
    \right]_{\xi=\xi_B},\\
    v_{0 M} &=& 2i \ell^3 \int^{\eta_0}_{-\infty}
    \frac{d\eta}{(\sinh \xi_B)^3 {\eta}^4}
    \phi_M^{(-)}(\eta, \xi_B) \left[
    \partial_{\xi} \tilde{\phi}_{0}^{(-)}(\eta, \xi)
    \right]_{\xi=\xi_B}.
    \label{v-coeffcient}
    \end{eqnarray}
To evaluate these coefficients, we need the source term $\left. \partial_{\xi} \tilde{\phi}_{0}^{(-)} \right|_{\xi=\xi_B}$ written in terms of the coordinates $(\eta, \xi)$, which is 
    \begin{eqnarray}
    \left. \partial_{\xi} \tilde{\phi}_{0}^{(-)}
    \right|_{\xi=\xi_B}
    = - \ell^{-3/2} \sinh \tilde{\xi}_B
    \frac{C(H_2)}{\sqrt{2p}}ip \epsilon_H
    \eta_0 \eta \sinh \xi_B
    e^{ip \sqrt{\eta^2+2\epsilon_H \eta_0 \eta \cosh \xi_B+ \epsilon_H^2 \eta_0^2}}.
    \label{boundary source}
    \end{eqnarray}

From Eqs.~(\ref{del phi}) and (\ref{del phi = mode sum}), 
we finally obtain the Bogoliubov coefficients relating the initial
zero mode or KK modes to the final zero mode, 
    \begin{eqnarray}
    \varphi = \sum_M \left(
    \alpha_{0 M} \phi_M^{(-)} +
     \beta_{0 M} \phi_M^{(+)}
     \right),
    \label{varphi:final form}
    \end{eqnarray}
where
    \begin{eqnarray}
    \alpha_{0 M} &=& U_{0 M} + u_{0 M},
    \nonumber
    \\
    \beta_{0 M} &=&  v_{0 M}.
    \label{alpha and beta}
    \end{eqnarray}
From these coefficients we can evaluate the number and the power spectrum of the generated gravitons.

\section{Evaluation of Bogoliubov Coefficients}

Now let us evaluate the expressions for the Bogoliubov coefficients 
obtained in the preceding section. 
We concentrate on the two limiting cases:
the low energy regime ($\ell H_1 \ll 1$) and 
high energy regime ($\ell H_1 \gg 1$).
We first evaluate the coefficients $\alpha_{00}$ and $\beta_{00}$, which 
relate the initial zero mode to the final zero mode. 
We keep the terms up to second
order in $\epsilon_H$ (or equivalently in $\delta H/H_1$).

Substituting Eq.~(\ref{boundary source}) into Eq.~(\ref{v-coeffcient}),
we have
    \begin{eqnarray}
    \beta_{00} = C(H_1)C(H_2)
    \frac{H_2}{H_1} \epsilon_H \eta_0
    \int^{\eta_0}_{-\infty} d\eta \left(
    \frac{1}{\eta^2} - \frac{i}{p \eta^3} \right)
    e^{-ip \left( \eta -
    \sqrt{\eta^2 + 2\epsilon_H \eta_0
    \eta \cosh \xi_B + \epsilon_H^2 \eta_0^2} \right)}.
    \end{eqnarray}
Because there is a factor $\epsilon_H$ in front of the integral, we
can neglect the correction of $O(\epsilon_H^2)$ in the integrand.
Then we can carry out the integration to obtain
    \begin{eqnarray}
    \beta_{00} \approx
    C(H_1)C(H_2) \frac{H_2}{H_1} \epsilon_H
    \frac{i}{2 p \eta_0} e^{- 2ip \eta_0
    - ip \eta_0 \epsilon_H \cosh \xi_B}.
    \end{eqnarray}
Similarly, we get 
    \begin{eqnarray}
    \alpha_{00} \approx
    U_{0 0}
    + C(H_1) C( H_2 )
    \frac{H_2}{H_1} \epsilon_H \left( 1 + \frac{i}{2 p \eta_0} \right)
    e^{- ip \eta_0 \epsilon_H \cosh \xi_B},
    \end{eqnarray}
where $U_{00}$ is given by Eq.~(\ref{U00}).

At low energies ($\ell H_1 \ll 1$), the Bogoliubov coefficients
$\alpha_{00}$ and $\beta_{00}$ become 
    \begin{eqnarray}
    \alpha_{00} &\approx&
    \left[
    1 + \frac{i}{2 p \eta_0}\frac{\delta H}{H_1}
    \right]
    e^{-ip \eta_0 \delta H / H_2},
    \\
    \beta_{00} &\approx& \frac{i}{2p \eta_0}
    \frac{\delta H}{H_1}
    e^{- 2ip \eta_0 -i p \eta_0 \delta H / H_2}.
    \label{zerozero}
    \end{eqnarray}
It is worth noting that these expressions are correct up to the second order in $\delta H/H_1$. 
This result agrees with the result of the four-dimensional calculation
$\alpha^{({\rm 4D})}$ and
$\beta^{({\rm 4D})}$.

At high energies ($\ell H_1 \gg 1$), the coefficients are
    \begin{eqnarray}
    \alpha_{00} &\approx&
    \left[
    1 + \frac{3}{2}
  \left(  \frac{i}{2 p \eta_0}\frac{\delta H}{H_1}  \right)
    + \frac{3}{8}\left( \frac{\delta H}{H_1}
    \right)^2
    - \frac{i p \eta_0}{2} \frac{\delta H}{H_1}
    - \frac{(p \eta_0)^2}{2}
    \left( \frac{\delta H}{H_1} \right)^2
    \right]
    e^{-ip \eta_0 \frac{\delta H}{H_1} - i p \eta_0 \frac{3}{2}
    \left( \frac{\delta H}{H_1} \right)^2},
    \label{alpha_00_high_energy}
\\
    \beta_{00}
    &\approx& \frac{3}{2}
    \left(  \frac{i}{2p \eta_0} \frac{\delta H}{H_1} \right)
    e^{-2ip \eta_0 -ip \eta_0  \frac{\delta H}{H_1}
     - i p \eta_0 \frac{3}{2}
    \left( \frac{\delta H}{H_1} \right)^2}.
    \end{eqnarray}
Here we stress that the last two terms in the 
square brackets of Eq.~(\ref{alpha_00_high_energy}),
both coming from $U_{00}$, 
are enhanced at $p|\eta_0| \gg 1$.

Next, we calculate the Bogoliubov coefficients $\alpha_{0 \kappa}$ and
$\beta_{0 \kappa}$, which relate the initial KK
modes to the final zero mode, up to the leading 
first order in $\delta H / H_1$.
Although we will calculate the power spectrum up to second order in
$\delta H/ H_1$ in the succeeding section, 
the expressions up to first order are sufficient 
for $\alpha_{0 \kappa}$ and $\beta_{0 \kappa}$, 
in contrast to the case for $\alpha_{00}$ and $\beta_{00}$.
Again, substituting $\partial_{\xi} \tilde{\phi}_{0}^{(-)} (\eta,
\xi_B)$ into Eq.~(\ref{v-coeffcient}), we have
    \begin{eqnarray}
    \beta_{0 \kappa} =
    \sqrt{2p} ~\ell^{3/2} C(H_2)
    \chi^*_{\kappa} (\xi_B)
    \frac{\sinh \tilde{\xi}_B}{(\sinh \xi_B)^2}
    \epsilon_H \eta_0 \int^{\eta_0}_{-\infty} \frac{d\eta}{\eta^3}
    \psi_{\kappa}^{(-)} (\eta) e
    ^{ip \sqrt{\eta^2 + 2 \epsilon_H \eta_0
    \eta \cosh \xi_B + \epsilon_H^2 \eta_0^2}}.
    \end{eqnarray}
Setting $\epsilon_H$ in the integrand to zero, the coefficient
reduces to
    \begin{eqnarray}
    \beta_{0 \kappa} \approx
    \sqrt{\frac{\pi}{2}} C( H_2 )
    \chi_{\kappa}^* ( \xi_B )
    \frac{\ell H_2}{(\ell H_1)^2}
    \epsilon_H p \eta_0
    e^{-\pi \kappa /2} \int^{p |\eta_0|}_{\infty}
    dx x^{-3/2} H^{(1)}_{i \kappa}(x) e^{i x},
    \label{beta_0k_int_wrt_x}
    \end{eqnarray}
where we have introduced the integration variable $x = -p \eta$.
Similarly we have
    \begin{eqnarray}
    \alpha^*_{0 \kappa} \approx U^*_{0 \kappa} -
    \sqrt{\frac{\pi}{2}} C( H_2 )
    \chi_{\kappa}^* ( \xi_B )
    \frac{\ell H_2}{(\ell H_1)^2}
    \epsilon_H p \eta_0
    e^{-\pi \kappa /2} \int^{p |\eta_0|}_{\infty}
    dx x^{-3/2} H^{(1)}_{i \kappa}(x) e^{-i x},
    \label{alpha_0k_int_wrt_x}
    \end{eqnarray}
where $U_{0 \kappa}$ is given by Eq.~(\ref{U0kappa}) and is ${\cal
O}(\epsilon_H)$\footnote{
The integrals including the Hankel function is written in terms of generalized hypergeometric functions.  
}.

Now let us discuss the dependence of $\alpha_{0 \kappa}$ and $\beta_{0
\kappa}$ on $\ell H_1$ and $\delta H/H_1$ in the limiting cases, $\ell
H_1 \ll 1$ and $\ell H_1 \gg 1$.  At low energies, we find, using
Eq.~(\ref{H_LL_1_and_kH_LL_1}), 
    \begin{eqnarray}
    |\beta_{0 \kappa}|^2,~|\alpha_{0 \kappa}|^2
    \propto   (\ell H_1)^2 \left( \frac{\delta H}{H_1} \right)^2,
\quad (\ell H_1 \ll 1) \,,
    \label{0k_in_low_case}
    \end{eqnarray}
where we have omitted the dependence on $\kappa$ and $p|\eta_0|$.
These coefficients are suppressed by the factors of $\ell H_1$ and $\delta H/H_1$.
Recall that $\alpha_{00}$ and $\beta_{00}$ agree with the standard
four-dimensional result at low energies. 
Thus, because of the suppression of $\alpha_{0 \kappa}$ and $\beta_{0
\kappa}$ at low energies, the four-dimensional result is recovered only
by the contribution from the initial zero mode.

At high energies, we obtain from Eq.~(\ref{H_GG_1_or_kH_GG_1})
    \begin{eqnarray}
    |\beta_{0 \kappa}|^2,~|\alpha_{0 \kappa}|^2
    \propto    \left( \frac{\delta H}{H_1} \right)^2 , 
    \quad (\ell H_1 \gg 1) \,,
    \label{independent of lH}
    \end{eqnarray}
where we have again omitted the dependence on $\kappa$ and $p|\eta_0|$.
In contrast to the result in the low energy regime (\ref{0k_in_low_case}),  
this high energy behavior is not associated with any suppression
factor. 

Integrating $|\beta_{0 \kappa}|^2$ over the KK continuum, we obtain the
total number of zero-mode gravitons created from the initial KK vacuum.
It can be shown that the coefficients behave 
as
$\beta_{0 \kappa} \approx - \alpha_{0 \kappa}^* \sim (p|\eta_0|)^{1/2}$
at $p|\eta_0| \ll 1$, and we have 
    \begin{eqnarray}
    \beta_{0 \kappa} + \alpha_{0 \kappa}^*
    \sim {\cal O} (p|\eta_0|)^{3/2}.
    \label{3/2jou}
    \end{eqnarray}
Thus the number of gravitons created is proportional to $p$ outside the horizon and is evaluated as
    \begin{eqnarray}
    \int^{\infty}_{0} | \beta_{0 \kappa}|^2
    d\kappa \approx
    \left\{ {\displaystyle
    0.5 \times p |\eta_0| (\ell H_1)^2
    \left( \frac{\delta H}{H_1} \right)^2,
    ~~~~(\ell H_1 \ll 1)},
    \atop {\displaystyle
    0.3 \times p |\eta_0|
    \left( \frac{\delta H}{H_1} \right)^2,
    ~~~~(\ell H_1 \gg 1)}.
    \right. 
    \label{beta_00 at p<<1}
    \end{eqnarray}
On the other hand, making use of the asymptotic form of the Hankel function
$H_{\nu}^{(1)}(x) \sim  e^{i(x-(2\nu+1)\pi/4)} /\sqrt{x}$ for $x \to
\infty$, we can evaluate the integral in Eq.~(\ref{beta_0k_int_wrt_x})
in the $p|\eta_0| \to \infty$ limit as
    \begin{eqnarray}
    \beta_{0 \kappa} \propto
    p|\eta_0| \int^{p|\eta_0|}_{\infty}
    dx \frac{e^{2ix}}{x^{2}}
    \sim \frac{1}{p|\eta_0|}  
    \quad  (p|\eta_0| \to \infty), 
    \label{decreasing_beta0k}
    \end{eqnarray}
where we have carried out the integration by parts and kept the most 
dominant term.
This shows that the creation of gravitons is suppressed well inside the
horizon. 
Since the 
assumption of the instantaneous transition tends to overestimate 
particle production at large $p$~\cite{Allen},  
the number of particles created from the initial zero mode and KK modes
is expected to be more suppressed inside the horizon than
Eq.~(\ref{decreasing_beta0k}) if we consider a realistic situation in
which the Hubble parameter changes smoothly. 
Note that $u_{0 \kappa} \propto p|\eta_0|
\int^{p|\eta_0|}_{\infty}x^{-2}dx$ is constant for $p|\eta_0| \to
\infty$. Therefore, the $\alpha_{0 \kappa}$ coefficient is dominated
by $U_{0 \kappa}$  at large $p$, which behaves like $|\alpha_{0 \kappa}|^2
\sim |U_{0 \kappa}|^2 \propto (p\eta_0)^2$.

\begin{figure}[htb]
  \begin{center}
   \includegraphics[keepaspectratio=true,height=50mm]{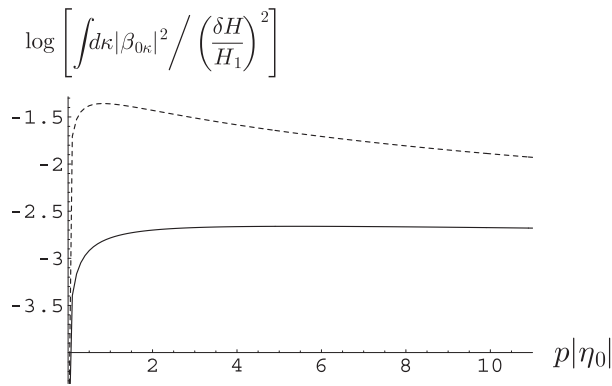}
  \end{center}
  \caption{Spectra of number density of zero-mode gravitons created from vacuum fluctuations in the initial KK modes.
  Integration over $\kappa$ is performed numerically.
  Solid line represents the low energy case with $\ell H_1 = 0.1$, while dashed line shows the high energy case. The latter should be understood as the limiting case $\ell H_1 \to \infty$ since at high energies the number does not depend on $\ell H_1$; see Eq.~(\ref{independent of lH}).}
  \label{fig:bogoliubov.eps}
\end{figure}

An example of numerical calculation is shown in Fig.~\ref{fig:bogoliubov.eps}. The figure shows that 
the spectrum has a peak around the Hubble scale and then decreases
inside the horizon, and we confirmed that the behavior of the number
density outside the horizon is well described by Eq.~(\ref{beta_00 at
p<<1}).

\section{Power Spectrum of Generated Gravitational Waves}

So far, we have discussed the Bogoliubov coefficients to see the number
of created gravitons.
However, our main interest is in the power spectrum
of gravitational waves 
because the meaning of ``particle'' is obscure at the super Hubble scale.

\subsection{Pure de Sitter brane}

Gravitational waves generated from pure de Sitter inflation 
on a brane have the scale invariant spectrum
    \begin{eqnarray}
    {\cal P}_{{\rm 5D}} = \frac{2C^2(H)}{M^2_{\mathrm{Pl}}}
    \left( \frac{H}{2\pi} \right)^2,
    \label{P_5D}
    \end{eqnarray}
which is defined by the expectation value of the 
squared amplitude of the vacuum fluctuation,
$ 8\pi p^3 \vert\phi_0\vert^2 /(2\pi M_5)^3$,
evaluated at a late time; see Eq.~(\ref{zeromode}). 
Since $C^2 \sim 1$ at $\ell H \ll 1$, this power spectrum 
agrees at low energies with the standard four-dimensional spectrum
    \begin{eqnarray}
    {\cal P}_{{\rm 4D}} = \frac{2}{M_{{\rm Pl}}^2}
    \left( \frac{H}{2\pi} \right)^2.
    \label{P_4D}
    \end{eqnarray}
At high energies, however, the power spectrum (\ref{P_5D}) is
enhanced due to the factor $C(H)$, and is much greater
than the four-dimensional counterpart.
This amplification effect was found in Ref.~\cite{Lan1}.
These results say that the difference between
Eq.~(\ref{P_5D}) and Eq.~(\ref{P_4D}) is absorbed by the
transformation
\begin{eqnarray}
    H \mapsto H C(H) .
    \label{H->HC}    
\end{eqnarray}

\subsection{Model with variation of Hubble parameter}

Now let us turn to the case in which the Hubble parameter is not constant. 
Time variation of the Hubble parameter during inflation brings a small
modification to the spectrum, and the resulting spectrum depends on
the wavelength. 
Here we consider the amplitude of vacuum fluctuation of the zero mode 
in the final state. 
It will be a relevant observable for the observers on
the brane at a late epoch because  
the KK mode fluctuations at super Hubble scale 
rapidly decay in the expanding universe due to 
its four-dimensional effective mass. 
Since the behavior of the zero-mode wave function 
at the infinite future $\tilde{\eta} \to 0$ is known from
Eq.~(\ref{zeromode}), we find that 
the vacuum fluctuation for zero mode at a late epoch is given by 
    \begin{eqnarray}
  \lim_{\tilde\eta \to 0}\left| \alpha_{0 0} \tilde{\phi}_0^{(+)}
    - \beta_{0 0}^* \tilde{\phi}_0^{(-)}
    \right|^2 + \int d\kappa \left|
    \alpha_{0 \kappa} \tilde{\phi}_0^{(+)}
    - \beta_{0 \kappa}^* \tilde{\phi}_0^{(-)}
    \right|^2
 =    \frac{C^2(H_2)}{\ell}\frac{H_2^2}{2p^3}
    \left(
    \left| \alpha_{00} + \beta_{00}^* \right|^2 +
    \int d\kappa \left| \alpha_{0 \kappa}
    + \beta_{0 \kappa}^* \right|^2 \right).
    \label{vac_exp}
    \end{eqnarray}
Multiplying this by $(2 / M_5^3) (p^3 / 2\pi^2)$ and recalling the relation $\ell M_5^3 = M^2_{{\rm Pl}}$, we obtain the power spectrum
    \begin{eqnarray}
    {\cal P}_{{\rm 5D}}(p) &=& {\cal P}_{{\rm 5D}}^{{\rm zero}}(p)
    + {\cal P}_{{\rm 5D}}^{{\rm KK}}(p),
\end{eqnarray}
with 
\begin{eqnarray}
    {\cal P}_{{\rm 5D}}^{{\rm zero}}(p)
    &=& \frac{2C^2(H_2)}{M^2_{\mathrm{Pl}}}
    \left( \frac{H_2}{2\pi} \right)^2
    |\alpha_{00} + \beta^*_{00}|^2,
    \cr
    {\cal P}_{{\rm 5D}}^{{\rm KK}}(p)
    &=& \frac{2C^2(H_2)}{ M^2_{\mathrm{Pl}} }
    \left( \frac{H_2}{2\pi} \right)^2
    \int^{\infty}_0 d \kappa
    |\alpha_{0 \kappa} + \beta^*_{0 \kappa}|^2.
    \end{eqnarray}
The power spectrum in the four-dimensional theory, computed in the same way, is given by
    \begin{eqnarray}
    {\cal P}_{{\rm 4D}} (p) =
    \frac{2}{M_{{\rm Pl}}^2}
    \left( \frac{H_2}{2 \pi} \right)^2
    \left|
    \alpha^{({\rm 4D})} + \beta^{*({\rm 4D})}
    \right|^2.
    \label{spectrum_4D}
    \end{eqnarray}
The appearance of the coefficient $\alpha$ in the power spectrum may 
look unusual.
This is due to our setup in which the final state of the 
universe is still inflating. In such a case, 
the fluctuations that have left the Hubble horizon never reenter it.
Outside the horizon, the number of particle created, does not
correspond to the power spectrum.

There are two apparent differences between ${\cal P}_{{\rm 5D}} $ and ${\cal P}_{{\rm 4D}} $; the normalization factor $C(H)$ and the contribution from the KK modes ${\cal P}_{{\rm 5D}}^{{\rm KK}} $.
In the low energy regime, however, the two spectra agree with each other:
\begin{eqnarray}
{\cal P}_{{\rm 5D}} \approx {\cal P}_{{\rm 4D}}
 \quad  (\ell H_1 \ll 1 ). 
\end{eqnarray}
This is because, as is seen from the discussion about 
the Bogoliubov coefficients in the preceding section, the zero-mode
contribution ${\cal P}_{{\rm 5D}}^{{\rm zero}} $ is exactly the same
as ${\cal P}_{{\rm 4D}} $ (up to the normalization factor), and the
Kaluza-Klein contribution ${\cal P}_{{\rm 5D}}^{{\rm KK}} $ is
suppressed by the factor $(\ell H_1)^2$.
On the other hand, when $\ell H_1$ is large, the amplitude of
gravitational waves deviates from the four-dimensional one owing to the
amplification of the factor $C(H)$. 

We have observed for the pure de Sitter inflation that the
correspondence between the five-dimensional power spectrum and the
four-dimensional one is realized by the map (\ref{H->HC}). 
It is interesting to investigate whether the correspondence can be
generalized to the present case.
It seems natural to give the transformation in this case by
    \begin{eqnarray}
    h \mapsto h C(h),
    \label{transformation:h->C(h)}
    \end{eqnarray}
where
    \begin{eqnarray}
    h(p) = H_2 \left|
    \alpha^{({\rm 4D})} + \beta^{*({\rm 4D})}
    \right|;
    \end{eqnarray}
namely, the rescaled power spectrum ${\cal P}_{{\rm res}}(p)$ is defined
as 
    \begin{eqnarray}
    {\cal P}_{{\rm res}}(p) =
    \frac{2 C^2(h)}{M_{{\rm Pl}}^2}
    \left( \frac{h}{2 \pi} \right)^2.
    \end{eqnarray}
We will see that this transformation works well and mostly absorbs the
difference between the five- and four- dimensional amplitudes.

We examine the differences between the five-dimensional spectrum and
the rescaled four-dimensional 
spectrum by expanding them with respect to $\delta H/H_1$ as 
    \begin{eqnarray}
    {\cal P}_{{\rm 5D}} (p)
    &=&
      {\cal P}_{{\rm 5D}}^{(0)}
    +  {\cal P}_{{\rm 5D}}^{(1)}
    + {\cal P}_{{\rm 5D}}^{(2)}  
    + {\cal O} \left( \frac{\delta H}{H}\right)^3
    + {\cal P}_{{\rm 5D}}^{{\rm KK}},
    \label{tenkai_S_5D}
    \\
    {\cal P}_{{\rm res}} (p)
    &=&
      {\cal P}_{{\rm res}}^{(0)} 
    + {\cal P}_{{\rm res}}^{(1)} 
    + {\cal P}_{{\rm res}}^{(2)} 
    + {\cal O} \left( \frac{\delta H}{H}\right)^3 \,.
    \label{tenkai_S_res}
    \end{eqnarray}
Here the quantity associated with the superscript $(n)$ represents 
the collection of the terms of ${\cal O} \left(({\delta H}/{H})^n\right)$. 
On the right hand side of Eq.~(\ref{tenkai_S_5D}), all the terms except
for the last one come from the initial zero mode.  The direct expansion
shows that the leading terms in Eqs.~(\ref{tenkai_S_5D}) and
(\ref{tenkai_S_res}) exactly agree with each other up to the first
order in $\delta H/H_1$, 
\begin{eqnarray}
         {\cal P}_{{\rm 5D}}^{(0)}
    +  {\cal P}_{{\rm 5D}}^{(1)}
    =        {\cal P}_{{\rm res}}^{(0)} 
    + {\cal P}_{{\rm res}}^{(1)} 
    =
\frac{2 C^2 ( H_1 )}{ M^2_{\mathrm{Pl}} }
    \left( \frac{H_1}{2 \pi} \right)^2
    \left\{ 1 + \left[ \frac{\sin( 2p \eta_0 )}{p \eta_0}
    -2 \right] \frac{C^2(H_1)}{\sqrt{1+ (\ell H_1)^2}}
    \frac{\delta H}{H_1} \right\} \,.
\end{eqnarray}
These terms contain only the contribution from the initial zero mode.

The contribution from the initial KK modes ${\cal P}_{{\rm 5D}}^{{\rm
KK}}$ is of order $(\delta H/H_1)^2$ because both $\alpha_{0 \kappa}$
and $\beta_{0 \kappa}$ are ${\cal O}(\delta H/H_1)$. Thus, to examine 
whether the agreement of the spectrum continues to hold 
even after including the KK modes, 
we investigate the second order part of the spectrum. 
The second order terms ${\cal P}_{{\rm 5D}}^{(2)}$ and ${\cal P}_{{\rm
res}}^{(2)}$ are given by 
\begin{eqnarray}
    {\cal P}_{{\rm 5D}}^{(2)}(p) 
    &=&
    \frac{2 C^2 ( H_1 )}{M^2_{\mathrm{Pl}}}
    \left( \frac{H_1}{2 \pi} \right)^2
    \Bigg[ ( p \eta_0 )^2 \left(
    \frac{C^{-2}( H_1 ) + C^2( H_1 )}
    {\sqrt{1 + (\ell H_1)^2}}
    - 2 \right) + \cos( p \eta_0) \left(
    \frac{C^2( H_1 )}{\sqrt{1 + (\ell H_1)^2}} +
    1 \right)
    \nonumber
    \\ &&
    + \frac{\sin^2 ( p \eta_0 )}{(p \eta_0)^2}
    \frac{C^2( H_1 )}{\sqrt{1 + (\ell H_1)^2}}
    + \frac{\sin (2 p \eta_0)}{2 p \eta_0}
    \left(
    \frac{2 + 3 (\ell H_1)^2}
    {1 + (\ell H_1)^2} -
    \frac{6 C^2( H_1 )}{\sqrt{1 + (\ell H_1)^2}}
    \right)
    \nonumber\\ &&
    + \frac{4 C^2( H_1 )}{\sqrt{1 + (\ell H_1)^2}}
    - \frac{3 + 4 (\ell H_1)^2}
    {1 + (\ell H_1)^2}
    \Bigg]
    \frac{C^2(H_1 )}{\sqrt{1 + (\ell H_1)^2}}
    \left(
    \frac{\delta H}{H_1}
    \right)^2,
    \label{P5D_2ndO}
    \\
    {\cal P}_{{\rm res}}^{(2)}(p)
    &=&
    \frac{2 C^2 ( H_1 )}{ M^2_{\mathrm{Pl}} }
    \left( \frac{H_1}{2 \pi} \right)^2
    \Bigg[ 2 \cos( p \eta_0)
    + \frac{\sin^2 ( p \eta_0)}{(p \eta_0)^2} +
    \frac{\sin (2 p \eta_0)}{p \eta_0}
    \left(
    \frac{2+ 3(\ell H_1)^2}
    {1 + (\ell H_1)^2} -
    \frac{4 C^2( H_1 )}{\sqrt{1 + (\ell H_1)^2}}
    \right)
    \nonumber\\ &&
    - \frac{\sin^2(2p\eta_0)}{(2p \eta_0)^2}
    \left(
    \frac{4+ 5(\ell H_1)^2}
    {1 + (\ell H_1)^2} -
    \frac{4 C^2( H_1 )}{\sqrt{1 + (\ell H_1)^2}}
    \right)
    \nonumber \\ &&
    + \frac{4 C^2( H_1 )}{\sqrt{1 + (\ell H_1)^2}}
    - \frac{3 + 4 (\ell H_1)^2}
    {1 + (\ell H_1)^2}
    \Bigg]
    \frac{C^2( H_1 )}{\sqrt{1 + (\ell H_1)^2}}
    \left(
    \frac{\delta H}{H_1}
    \right)^2.
\end{eqnarray}
From these equations, we notice that 
${\cal P}_{{\rm res}}^{(2)}$ and ${\cal
P}_{{\rm 5D}}^{(2)}$ do not agree with each other.  
We see that the difference is enhanced in particular at $p|\eta_0|\gg 1$. 
However, as we have mentioned earlier, the KK mode contribution
also gives a correction of the same order, and in fact we show that 
approximate agreement is recovered even in this order by adding the KK mode 
contribution.

First we observe the power spectrum at $p|\eta_0|\ll  1$. 
Expanding ${\cal P}_{{\rm 5D}}^{(2)}$ and ${\cal P}_{{\rm res}}^{(2)}$
with respect to $p|\eta_0|$, we have
    \begin{eqnarray}
    {{\cal P}_{{\rm 5D}}^{(2)} (p)\over {{\cal P}^{(0)}}}~,~
    {{\cal P}_{{\rm res}}^{(2)} (p)\over {{\cal P}^{(0)}}} \sim
    (p \eta_0)^2 \left( \frac{\delta H}{H_1} \right)^2. 
    \end{eqnarray}
On the other hand, Eq.~(\ref{3/2jou}) leads to 
    \begin{eqnarray}
    {{\cal P}_{{\rm 5D}}^{{\rm KK}} (p) \over {{\cal P}^{(0)}}}\sim
    (p |\eta_0|)^3 \left( \frac{\delta H}{H_1} \right)^2.
    \end{eqnarray}
Therefore the difference is small outside the horizon as 
\begin{eqnarray}
    \left| {{\cal P}_{{\rm 5D}}^{(2)}
    + {\cal P}_{{\rm 5D}}^{{\rm KK}}
    - {\cal P}_{{\rm res}}^{(2)}\over {{\cal P}^{(0)}}} \right|
    \sim (p\eta_0)^2 \left( \frac{\delta H}{H_1} \right)^2
    \quad (p|\eta_0| \ll 1) \,, 
\end{eqnarray}
although the cancellation between 
${\cal P}_{{\rm 5D}}^{(2)} (p)$ and ${\cal P}_{{\rm res}}^{(2)} (p)$ 
does not happen. 
By a similar argument, at $p|\eta_0|\lesssim 1$, we have 
\begin{eqnarray}
    \left| {{\cal P}_{{\rm 5D}}^{(2)}
    + {\cal P}_{{\rm 5D}}^{{\rm KK}}
    - {\cal P}_{{\rm res}}^{(2)}\over {{\cal P}^{(0)}}} \right|
      \sim \left( \frac{\delta H}{H_1} \right)^2
      \quad (p|\eta_0| \lesssim 1) \,.
\end{eqnarray}

The situation is more interesting when we consider the spectrum inside
the Hubble horizon.
There is a term proportional to $(p \eta_0)^2$ in
${\cal P}_{{\rm 5D}}^{(2)}$, which is dominant 
at $p|\eta_0| \gg 1$, while there is no corresponding 
term in ${\cal P}_{{\rm res}}^{(2)}$. 
Hence, the difference between 
${\cal P}_{{\rm 5D}}^{(2)}$ 
and ${\cal P}_{{\rm res}}^{(2)}$ is 
    \begin{eqnarray}
    {{\cal P}_{{\rm 5D}}^{(2)} - {\cal P}_{{\rm res}}^{(2)}\over 
      {{\cal P}^{(0)}}}
    \approx - 
    ( p \eta_0 )^2 \left(
    2- \frac{C^{-2}( H_1 ) + C^2( H_1 )}
    {\sqrt{1 + (\ell H_1)^2}}
    \right) \frac{C^2(H_1 )}{\sqrt{1 + (\ell H_1)^2}}
    \left(
    \frac{\delta H}{H_1}
    \right)^2.
    \label{p-p_inside}
    \end{eqnarray}
Our approximation is valid for $p|\eta_0| \delta
H/H_1 \lesssim 1$ (\ref{assumption 2}). 
Hence, within the region of validity, 
this difference can be as large as ${\cal P}^{(0)}$. 
We also note that the terms proportional to $(p|\eta_0|)^2$ 
come from $\alpha_{00}$, while the contribution from $\beta_{00}$ is 
suppressed at $p|\eta_0| \gg 1$. 
On the other hand, the contribution from initial KK modes, ${\cal
P}_{{\rm 5D}}^{{\rm KK}}(p)$, is dominated by $\alpha_{0 \kappa}$ at
$p|\eta_0| \gg 1$: ${\cal P}_{{\rm 5D}}^{{\rm KK}} \propto \int
d\kappa|\alpha_{0 \kappa} +\beta^*_{0 \kappa}|^2 \sim \int d\kappa
|\alpha_{0 \kappa}|^2$.
This means that, although the creation of zero-mode gravitons from the
initial KK modes is negligible inside the horizon, a part of the
amplitude of the final zero mode comes from the initial 
KK modes losing their KK momenta.
Since the coefficient $\alpha_{0 \kappa}$ 
is proportional to $p |\eta_0|$
at large $p|\eta_0|$, 
${\cal P}_{{\rm 5D}}^{{\rm KK}}$ behaves as
    \begin{eqnarray}
    {{\cal P}_{{\rm 5D}}^{{\rm KK}}(p)\over {{\cal P}^{(0)}}} \sim
    + (p\eta_0)^2 \left(
    \frac{\delta H}{H_1} \right)^2.
    \end{eqnarray}
This KK mode contribution cancels ${\cal P}^{(2)}_{\rm 5D}$. 
The cancellation can be proved by looking at the property of the Bogoliubov coefficients
    \begin{eqnarray}
    |\alpha_{00}|^2 - |\beta_{00}|^2
    + \int d\kappa (|\alpha_{0 \kappa}|^2
    -|\beta_{0 \kappa}|^2) =1. 
    \end{eqnarray}
This relation together with Eq.~(\ref{vac_exp}) implies that the power spectrum cannot significantly 
deviate from $C^2(H_2) H_2^2/2\ell p^3$ 
in the region where $\beta_{0 0}$ and $\beta_{0 \kappa}$ are negligibly small. 
We can also demonstrate the cancellation by explicit calculation  
in the low and high energy limits.
For $p|\eta_0| \gg 1$, we have $\alpha_{0 \kappa} \approx U_{0 \kappa}$. Then from  Eqs.~(\ref{H_LL_1_and_kH_LL_1}), (\ref{H_GG_1_or_kH_GG_1}), and (\ref{U0kappa}), we obtain 
    \begin{eqnarray}
    \int d\kappa |\alpha_{0 \kappa}|^2
    \approx \left\{
    {\displaystyle
    \int^{\infty}_{0} \frac{2\kappa \tanh (\pi \kappa) d\kappa}
    {(\kappa^2 +1/4) (\kappa^2+9/4)}
    \times (p \eta_0)^2 (\ell H_1)^2
    \left( \frac{\delta H}{H_1} \right)^2
    ~~~~~~(\ell H_1 \ll 1),
    }
    \atop {\displaystyle
    \frac{6}{\pi} \int^{\infty}_{0} \frac{\kappa^2 d\kappa}
    {(\kappa^2 +1/4)^2 (\kappa^2+9/4)}
    \times(p \eta_0)^2 \left( \frac{\delta H}{H_1} \right)^2
    ~~~~~~(\ell H_1 \gg 1),
    }\right. 
    \end{eqnarray}
which gives $(p\eta_0)^2 (\ell H_1)^2 (\delta H/H_1)^2$ in the low energy
regime and $(3/4) (p \eta_0)^2 (\delta H/H_1)^2$ in the high energy
regime.
Comparing these with Eq.~(\ref{p-p_inside}), we see that 
${\cal P}_{{\rm 5D}}^{{\rm KK}}(p)$ cancels 
${\cal P}_{{\rm 5D}}^{(2)}(p)$\footnote{The small $\ell H$ expansion $C^2(H) \approx 1 - (\ell
H)^2 [1/2 + \ln (\ell H/2)]$ is used here to investigate the low energy case.}.

\begin{figure}[htb]
  \begin{center}
   \includegraphics[keepaspectratio=true,height=65mm]{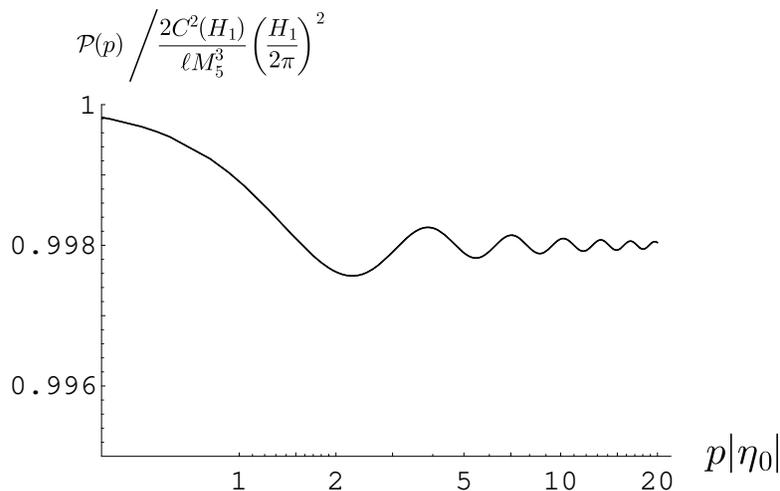}
  \end{center}
  \caption{Five-dimensional power spectrum of gravitational waves
  ${\cal P}_{{\rm 5D}}(p)$ and the four-dimensional one ${\cal
  P}_{{\rm 4D}}(p)$ at low energies ($\ell H_1 = 10^{-2}$) with
  $\delta H/H_1 = 10^{-3}$. These two agree 
  with each other. In this case initial KK modes give negligible contribution.}
  \label{fig:spectrum_for_LOW.eps}
\end{figure}

\begin{figure}[htb]
  \begin{center}
   \includegraphics[keepaspectratio=true,height=65mm]{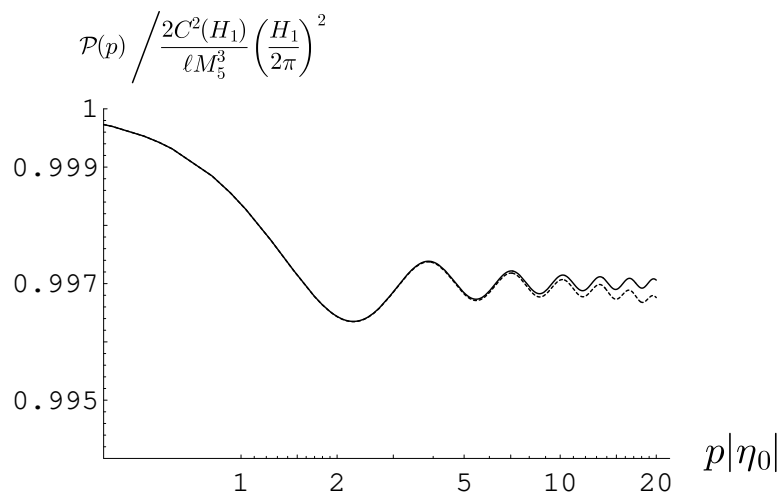}
  \end{center}
  \caption{Power spectra of gravitational waves at high energies
  ($\ell H_1 = 10^3$) with $\delta H/H_1 = 10^{-3}$.  Five-dimensional
  spectrum ${\cal P}_{{\rm 5D}}^{{\rm zero}}(p)+{\cal P}_{{\rm
  5D}}^{{\rm KK}}(p)$ and the rescaled four-dimensional one ${\cal
  P}_{{\rm res}}(p)$ agree well with each other (solid lines), while only the
  zero-mode contribution ${\cal P}_{{\rm 5D}}^{{\rm zero}}(p)$
  gives the reduced fluctuation amplitude inside the horizon (dotted line).}
  \label{fig:spectrum_for_HIGH.eps}
\end{figure}

To summarize, we have observed that the agreement between the rescaled
spectrum ${\cal P}_{{\rm res}}(p)$ and the five-dimensional spectrum
${\cal P}_{{\rm 5D}}(p)$ is exact up to first order in $\delta
H/H_1$. The agreement is not exact at second order, but 
we found that the correction is not enhanced at any wavelength 
irrespective of the value of $\ell H_1$. 
Just for illustrative purpose   
we show the results of numerical calculations in
Fig.~\ref{fig:spectrum_for_LOW.eps} and
Fig.~\ref{fig:spectrum_for_HIGH.eps}.


\section{Discussion}

In this paper we have investigated the generation of primordial 
gravitational waves and its power spectrum in the inflationary 
braneworld model, focusing on the effects of the variation of the Hubble 
parameter during inflation. 
For this purpose, we considered a model in which 
the Hubble parameter changes discontinuously. 

In the case of de Sitter inflation with constant Hubble parameter
$H$, the spectrum is known to be given by Eq.~(\ref{P_5D})
\cite{Lan1}. It agrees with the standard four-dimensional one
(Eq.~(\ref{P_4D})) at low energies $\ell H \ll 1$, but at high energies
$\ell H \gg 1$ it significantly deviates from Eq.~(\ref{P_4D}) due to the
amplification effect of the zero-mode normalization factor $C(H)$. 
One can say, however, the five-dimensional spectrum is obtained from the
four-dimensional one by the map $H \mapsto HC(H)$.

In a model with variable Hubble parameter, 
gravitational wave perturbations are expected to be generated 
not only from the ``in-vacuum'' of the zero mode but also 
from that of the Kaluza-Klein modes. 
Hence, it is not clear whether there is a simple relation 
between the five-dimensional spectrum and the four-dimensional
counterpart. 
Analyzing the model with a discontinuous jump in the Hubble parameter, 
we have shown that this is indeed approximately the case.
More precisely, if the squared amplitude of four-dimensional
fluctuations is given by $(h/2\pi M_{\rm Pl})^2$, 
we transform $(h/2\pi M_{\rm Pl})^2$ into
$C^2(h) (h/2\pi M_{\rm Pl})^2$, then the resulting rescaled spectrum 
exactly agrees with the five-dimensional spectrum ${\cal P}_{{\rm 5D}}(p)$ up
to first order in $\delta H /H_1$. 
At second order ${\cal O}(\delta H/H)^2$ 
the agreement is not exact, but the difference is not enhanced 
at any wavelength irrespective of the value of $\ell H_1$. 
Hence, in total, the agreement is not significantly disturbed by the 
mismatch at second order.  
As a non-trivial point, 
we also found that the initial KK mode vacuum fluctuations can give
non-negligible contribution to the final zero-mode states at second order.

One may expect that the power spectrum of gravitational waves 
in the braneworld model would reflect the characteristic length scale 
corresponding to the curvature (or ``compactification'') scale of
the extra dimension $\ell$. 
However, our analysis showed that the resultant power spectrum does not
depend on the ratio of the wavelength of gravitational waves to the bulk
curvature scale, $p|\eta_0| \ell H$.

Here we should mention the result of Ref.~\cite{Rubakov} that is summarized in Appendix \ref{app_Ruba}.
Their setup is the most violent version of the transition $H_1 \to H_2 = 0$.
If the wavelength of the gravitational waves is much longer than both
the Hubble scale and the bulk curvature scale, the power
spectrum is given by Eq.~(\ref{VF_dS->Mink}) and obviously it is
obtained from the four-dimensional counterpart by the map $H \mapsto HC(H)$.
However, if the wavelength is longer than the Hubble scale but much
smaller than the bulk curvature radius, the amplitude is highly damped
as is seen from Eq.~(\ref{damped amplitude}) and the map $H \mapsto
HC(H)$ does not work at all. 
This damping of the amplitude can be understood in the following way.
In the high energy regime $\ell H \gg 1$, the motion of the
brane with respect to the static bulk is ultrarelativistic. 
At the moment of transition to the Minkowski phase, the brane abruptly stops. 
Zero-mode gravitons with wavelength smaller than the bulk curvature
scale can be interpreted as ``particles'' traveling in the five 
dimensions. These gravitons make a ``hard hit'' with the brane 
at the moment of this transition, and get large momenta in the fifth direction 
relative to the static brane. As a result, these gravitons 
escape into the bulk as KK gravitons, and 
thus the amplitude (\ref{damped amplitude}) is damped.
On the other hand, in our model the change of the Hubble parameter is assumed to
be small, and hence such violent emission of KK gravitons does not happen. 
If this interpretation is correct, 
the mapping rule $h \mapsto hC(h)$ will generally give a rather 
good estimate for the prediction of inflationary braneworld models 
as far as time variation of the Hubble parameter 
is smooth. 
If we can confirm the validity of this prescription in more general 
cases, the analysis of gravitational wave perturbations will be 
simplified a lot. 
We would like to return to this issue of generalization 
in a future publication.

\acknowledgments
H.K. is supported by the JSPS. To complete this work, the discussion during and 
after the YITP workshop YITP-W-02-19 was useful. 
This work is partly supported by the Monbukagakusho
Grant-in-Aid No.~1470165~(T.T.).

            \appendix

\section{Particle creation when connected to Minkowski brane \label{app_Ruba}}
Here for comparison with our results 
we briefly summarize the results obtained by 
Gorbunov {\it et al.}~\cite{Rubakov} 
focusing on the power spectrum of gravitational waves. 
Their method is basically the same that we have already explained 
in the main text. 
They considered the situation that de 
Sitter inflation on the brane with constant Hubble parameter $H$ 
suddenly terminates at a conformal time $\eta =\eta_0$, 
and is followed by a Minkowski phase. 
The power spectrum of gravitational waves is expressed in terms of the
Bogoliubov coefficients as
    \begin{eqnarray}
    {\cal P}_{{\rm 5D}} (p) =
    \frac{2}{M^2_{{\rm Pl}}}
    \left( \frac{H}{2 \pi} \right)^2
    (p \eta_0)^2 \left(
    1 + 2|\beta_{00}|^2 + 2 \int d \kappa
    |\beta_{0 \kappa}|^2 \right),
    \end{eqnarray}
where $|\beta_{00}|^2$ and $|\beta_{0 \kappa}|^2$ are the number of
zero-mode gravitons created from initial zero-mode and KK modes,
respectively.
At super Hubble scale ($p|\eta_0| \ll 1$) we can neglect the
first term in the parentheses, which corresponds to the vacuum 
fluctuations in Minkowski space.

According to \cite{Rubakov}, when $p |\eta_0| \ell H \ll 1$ 
(i.e., the wavelength of gravitational wave
$(p/a)^{-1}$ is much larger than the bulk curvature scale $\ell$) and
$p|\eta_0| \ll 1$, the 
coefficients are given by
    \begin{eqnarray}
    |\beta_{00}|^2 &\approx& \frac{C^2(H)}{4(p\eta_0)^2},
    \\
    \int d\kappa |\beta_{0 \kappa}|^2
    &\sim& \left\{
    p|\eta_0| (\ell H)^2
    ~~~~~~(\ell H \ll 1),
    \atop
    p|\eta_0| \ell H
    ~~~~~~(\ell H \gg 1).
    \right. 
    \end{eqnarray}
One can see that the contribution from initial KK modes is suppressed 
irrespective of the expansion rate $\ell H$.
Therefore the power spectrum is evaluated as
    \begin{eqnarray}
    {\cal P}_{{\rm 5D}} \approx
    \frac{C^2(H)}{M^2_{{\rm Pl}}}
    \left( \frac{H}{2 \pi} \right)^2.
    \label{VF_dS->Mink}
    \end{eqnarray}
Equation~(\ref{VF_dS->Mink}) is half of the power spectrum on the de Sitter
brane (Eq.~(\ref{P_5D})), 
and this result can be understood as follows.
The amplitude of fluctuations at the super Hubble scale stays constant. 
After the sudden transition from the de Sitter phase to
the Minkowski phase, the Hubble scale becomes infinite. 
Therefore, those fluctuation modes are now inside the Hubble horizon, 
and they begin to oscillate. 
As a result, the mean-square vacuum fluctuation becomes half of the
initial value.

On the other hand, when  $p |\eta_0| \ell H \gg 1$ and $p|\eta_0| \ll
1$ (these conditions require $\ell H \gg 1$), the Bogoliubov
coefficients are given by 
    \begin{eqnarray}
    |\beta_{00}|^2 &\approx& \frac{C^2(H)}{(p\eta_0)^2}
    \frac{1}{(p\eta_0 \ell H)^2},
    \\
    \int d\kappa |\beta_{0 \kappa}|^2
    &\sim& \frac{1}{(p\eta_0 \ell H)^2}.
    \end{eqnarray}
As before, the contribution from initial KK modes is negligible, 
and that from $|\beta_{00}|^2$ dominates the power spectrum,
    \begin{eqnarray}
    {\cal P}_{{\rm 5D}} \approx
    \frac{C^2(H)}{M^2_{{\rm Pl}}}
    \left( \frac{H}{2 \pi} \right)^2
    \frac{4}{(p\eta_0 \ell H)^2}.
    \label{damped amplitude}
    \end{eqnarray}
One can see that the spectrum is suppressed by 
the factor $4/(p\eta_0 \ell H)^{2}$.

\section{Calculations of the inner product}
\label{app:inner pro}

We derive Eqs.~(\ref{U00}) and (\ref{U0kappa}) by calculating the inner product (\ref{innerp}).
Expanding Eq.~(\ref{tilde eta xi}) in terms of $\epsilon_H$, we have
    \begin{eqnarray}
    \tilde{\eta} &=& \eta +
    \epsilon_H \eta_0 \cosh \xi -\epsilon_H^2 \eta^2_0
    (\sinh \xi)^2 / (2 \eta) + \cdots,
    \label{tilde_AND_eta-relation}\\
    \tilde{\xi} &=& \xi - \epsilon_H \eta_0
    \sinh \xi / \eta
    + \epsilon_H^2 \eta^2_0 \cosh \xi \sinh \xi
    / \eta^2 + \cdots,
    \end{eqnarray}
which reduce to the following forms at the infinite past ($\eta \to - \infty$),
    \begin{eqnarray}
    \tilde{\eta} - \eta &=& \epsilon_H \eta_0 \cosh \xi
    \label{eta_inf1},\\
    \tilde{\xi} - \xi &=& 0.
    \label{eta_inf2}
    \end{eqnarray}
Then, $U_{00}$ is evaluated at $\eta = - \infty$ as 
    \begin{eqnarray}
    U_{0 0} &=& \left.
    - 2 i \ell^3 \int^{\infty}_{\xi_B}
    \frac{d \xi}{\eta^2 (\sinh \xi )^3}
    \left(
    \tilde{\phi}^{(-)}_0 \partial_{\eta}
    \phi^{(+)}_0 - \phi^{(+)}_0
    \partial_{\eta} \tilde{\phi}^{(-)}_0
    \right) \right|_{\eta \to - \infty}
    \nonumber\\
    &=& \frac{H_2}{H_1}
    C (H_1) C(H_2)
    \cdot 2 \ell^2 H_1^2 \int^{\infty}_{\xi_B}
    d \xi \frac{e^{- i \epsilon_H p \eta_0 \cosh \xi}}
    {(\sinh \xi)^3}
    \nonumber\\
    &\approx& \frac{H_2 C(H_2)}
    {H_1 C (H_1)} e^{
    - i\epsilon_H p \eta_0 \cosh \xi_B}
    \cdot 2 \ell^2 H_1^2 C^2(H_1) \times
    \nonumber\\
    &{}&\times e^{
    i\epsilon_H p \eta_0 \cosh \xi_B}
    \int^{\infty}_{\xi_B} d \xi
    \frac{1}{(\sinh \xi)^3}
    \left[ 1 - i \epsilon_H p \eta_0
    \cosh \xi - \frac{1}{2}( \epsilon_H p \eta_0 )^2
    ( \cosh \xi )^2 \right],
    \end{eqnarray}
where we expanded the integrand with respect to $\epsilon_H$ in the last line.
Integrating each term, we finally obtain
    \begin{eqnarray}
    U_{0 0} &\approx& \frac{H_2 C(H_2)}
    {H_1 C (H_1)} e^{
    - i\epsilon_H p \eta_0 \cosh \xi_B}
    \left[ 1 - i \epsilon_H p \eta_0
    \left(
    C^2 (H_1) - \cosh \xi_B
    \right)
    - \frac{1}{2}( \epsilon_H p \eta_0 )^2
    ( \sinh \xi_B )^2 \right].
    \end{eqnarray}
Note that the condition $\epsilon_H p |\eta_0| \cosh \xi_B (\approx p
|\eta_0| \delta H / H) \ll 1$ is required in order to justify 
the expansion of the exponent. 
Because the integral is saturated at $\xi \approx \xi_B$, we do not
have to worry about the validity of the expansion for large
$\cosh \xi$.

The explicit form of $U_{0 \kappa}$ is needed up to the first order in
$\epsilon_H$. 
A similar calculation leads to
    \begin{eqnarray}
    U_{0 \kappa}  &=& \left.
    - 2 i \ell^3 \int^{\infty}_{\xi_B}
    \frac{d \xi}{\eta^2 (\sinh \xi )^3}
    \left(
    \tilde{\phi}^{(-)}_0 \partial_{\eta}
    \phi^{(+)}_{\kappa} - \phi^{(+)}_{\kappa}
    \partial_{\eta} \tilde{\phi}^{(-)}_0
    \right) \right|_{\eta \to - \infty}
    \nonumber\\
    &=& - 2i \ell^3 \cdot \ell^{-1/2} C(H_2)
    \frac{H_2}{\sqrt{2p}} \cdot
    \frac{\ell^{-3/2}}{\sqrt{2p}}
    \int^{\infty}_{\xi_B}
    \frac{d \xi}{\eta^2 (\sinh \xi)^3}
    \chi_{\kappa} (\xi)
    \nonumber\\ &{}&\times
    \left. \left[ \left( \tilde{\eta}
    - \frac{i}{p} \right) e^{-ip \tilde{\eta}}
    \cdot (-1 - ip \eta) e^{ip \eta + i \pi/4}
    - \left( -\tilde{\eta} \right) e^{ip \eta + i \pi/4}
    \cdot (-ip \eta) e^{-ip \tilde{\eta}}
    \right] \right|_{\eta \to - \infty}
    \nonumber\\
    &=& - 2 e^{i \pi /4} \ell H_2
    C(H_2) \int^{\infty}_{\xi_B} d \xi
    \frac{e^{- i\epsilon_H p \eta_0 \cosh \xi_B}}
    {(\sinh \xi)^3} \chi_{\kappa} ( \xi).
    \end{eqnarray}
The integral in the last line, which we call $I$, can 
be calculated as follows. 
Again, expanding the integrand in terms of $\epsilon_H$, 
we have 
    \begin{eqnarray}
   I \approx \int^{\infty}_{\xi_B} d \xi
    \frac{\chi_{\kappa} ( \xi)}
    {(\sinh \xi)^3} \left(
    1 - i \epsilon_H p \eta_0 \cosh \xi
    \right).
    \label{U_0k_kinji}
    \end{eqnarray}
Let us consider the first term in the parentheses.
The spatial wave function $\chi_{\kappa}$ satisfies $(\sinh \xi)^{-3} \chi_{\kappa} = - (\kappa^2 +9/4)^{-1} \partial_{\xi}[(\sinh \xi)^{-3} \partial_{\xi} \chi_{\kappa}]$ (Eq.~(\ref{sep_space_KK})) with the boundary condition $\partial_{\xi} \chi_{\kappa}(\xi_B)=0$.
Therefore, together with the behavior at infinity, $(\sinh \xi)^{-3} \partial_{\xi} \chi_{\kappa} \sim (\sinh \xi)^{-3} \partial_{\xi} (\sinh \xi)^{3/2} \to 0$, we find that the integral of the first term vanishes. 
Then, using the integration by parts twice, we have
    \begin{eqnarray}
    \left( \kappa^2 + \frac{9}{4} \right)
    I &\approx& i \epsilon_H p \eta_0
    \int^{\infty}_{\xi_B} d \xi
    \cosh \xi
    \frac{\partial}{\partial \xi}
    \left[
    \frac{1}{(\sinh \xi)^3}
    \frac{\partial}{\partial \xi} \chi_{\kappa}
    ( \xi )
    \right] \nonumber\\
    &=& - i \epsilon_H p \eta_0
    \int^{\infty}_{\xi_B} d \xi
    \frac{1}{(\sinh \xi )^2}
    \frac{\partial}{\partial \xi} \chi_{\kappa} (\xi)
    \nonumber\\
    &=&  i \epsilon_H p \eta_0
    \frac{\chi_{\kappa} (\xi_B)}{(\sinh \xi_B )^2}
    + i \epsilon_H p \eta_0
    \int^{\infty}_{\xi_B} d \xi
    \frac{-2 \cosh \xi}{(\sinh \xi )^3}
    \chi_{\kappa} (\xi)
    \nonumber\\
    &\approx&
    i \epsilon_H p \eta_0
    \frac{\chi_{\kappa} (\xi_B)}{(\ell H_1)^2}
    + 2 I ,
    \end{eqnarray}
from which we can evaluate $U_{0 \kappa}$.
Note that the approximation is valid when $p |\eta_0| \delta H/H_1 \ll 1$.



\end{document}